# Towards Cross-layer Reliability Analysis of Transient and Permanent Faults

Hananeh Aliee†, Liang Chen‡, Mojtaba Ebrahimi‡, Michael Glaß†, Faramarz Khosravi†, and Mehdi B. Tahoori‡
†Friedrich-Alexander-Universität Erlangen-Nürnberg (FAU), Germany
‡Karlsruhe Institute of Technology (KIT), Germany

*Abstract*—Due to the increasing complexity of Multi-Processor Systems on Chip (MPSoCs), system-level design methodologies have got a lot of attention in recent years. However, the significant gap between the system-level reliability analysis and the level where the actual faults occur necessitates a cross-layer approach in which the sufficient data about the effects of faults at low levels are passed to the system level. So far, the cross-layer reliability analysis techniques focus on a specific type of faults, e. g., either permanent or transient faults. In this work, we aim at proposing a cross-layer reliability analysis which considers different fault types concurrently and connects reliability analysis techniques at different levels of abstraction using adapters.

## I. Introduction

Constantly shrinking device structures lead to produce and design smaller, more efficient, and often also less expensive system components. However, this results in higher susceptibilities to, e. g., environmental disturbances, manufacturing tolerances, and aging effects, and subsequently increase in the unreliability of the employed components [1]. Therefore, future design methodologies need to optimize and synthesize dependable embedded systems from unreliable components, not only for safety-critical, but for all types of application domains [2].

The complexity of Multi-Processor Systems on Chip (MP-SoCs) which are composed of many unreliable components is increasing almost daily. Due to this complexity, detailed low-level analysis approaches may not be applicable anymore and, as a result, (automatic) system-level design methodologies have gained a lot of attention. However, while system-level design approaches focus on the interplay of hardware and software, the actual faults occur on very low levels of abstraction. Therefore, a significant *gap* exists not only between the cause of faults and the fault-tolerant techniques to compensate them, but also between the reliability analysis techniques and their accuracy, e. g., what is the effect of a memory bit flip at gate level on the applications at system level.

Today, there exists a variety of reliability analysis techniques for both, the relatively low levels of abstraction that focus on technology as well as system-level analysis techniques that focus on the applications. Existing *cross-layer reliability analysis* techniques collect knowledge at lower levels of abstraction by combining different analysis techniques, and provide proper data for the analysis at higher levels of abstraction by performing abstraction and conversion [3]. So far,

Supported in part by the German Research Foundation (DFG) as part of the priority program *Dependable Embedded Systems* (SPP 1500).

these techniques focus on a specific type of faults, e. g., either *permanent* or *transient* faults. Given that different fault types affect the system differently and may also require different countermeasures, the outlined lack of suitable cross-layer analysis techniques that consider different fault types concurrently prohibits an adequate system-wide analysis during the design of MPSoCs.

This work proceeds towards a *Cross-layer Reliability Analysis* (CRA) as a combination of various reliability analysis techniques across different levels of abstraction [3], aiming at concurrent consideration of various fault types, e. g., permanent and transient faults. It intends to close the mentioned gaps by introducing the articulation points where the results of low-level component-wide reliability analysis techniques can be applied as design parameters to the system-level analysis approaches. The following points outline the key challenges one faces when tackling both permanent and transient fault types concurrently:

- The time scale of fault occurrence significantly differs from one fault type to the other.
- Once a component is found to be permanently defective, any transient reliability analysis on this component renders ineffective.
- The low-level reliability analysis techniques to analyze the effects of different fault types are not necessarily applied at the same level of abstraction.

## II. Problem Statement and Related Work

This work targets the system-level design of embedded MPSoCs that typically consist of several processor cores connected by a communication infrastructure such as shared on-chip buses, networks-on-a-chip, or field buses. We intend to propose a reliability analysis framework in which component-wide reliability analysis can be performed in details at lower levels where faults occur and a probabilistic error function is formulated for each component. Design decisions can be made later at higher levels of abstraction without any further concern about the low-level details (see Fig 1).

Currently, there are few works in the state-of-the-art which aim at providing a cross-layer approach for propagating the effects of faults at lower levels to higher levels of abstraction. The work in [4] proposes a cross-layer model named Resilience Articulation Point (RAP), providing a probabilistic fault abstraction and error propagation framework. This work shows that all physically induced faults at a CMOS device manifest as a single or multiple bit-flip(s) which can be detected or recovered at higher levels of abstraction. Therefore, the reliability analysis can be performed at higher levels of abstraction (e.g., system level) without concerning about the details of lower abstraction levels. A framework is provided by RAP to integrate different fault origins in an error function covering several physical faults as probabilistic bit-flips.

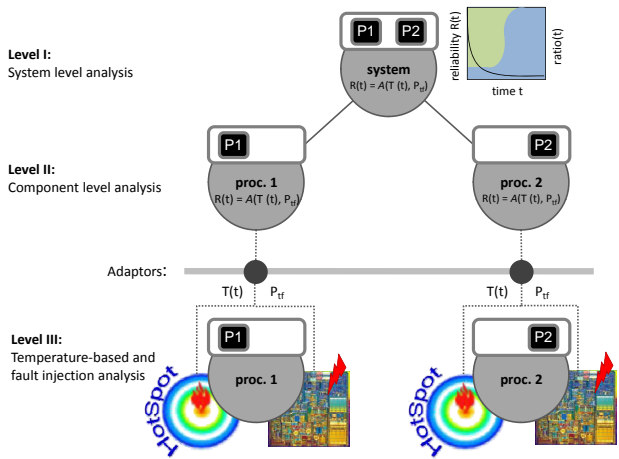

Fig. 1. An example of a cross-layer analysis that consists of three reliability levels of abstraction: At the highest, i.e., system level, the reliability of the system in the existence of both transient and permanent faults are derived in addition to the ratio of the reliability of these two fault types showing which fault type is more destructive to the system at each point of time. In the next level, the MPSoC is decomposed into subsystems, i.e., two processing units in this example. At level III, temperature analysis (for permanent faults) and fault injection analysis (for transient faults) are performed and passed back to level II. Finally, the individual reliability functions from level II are passed to the system-level.

The work in [3] focuses on reliability analysis and improvement in MPSoCs using a *Cross-layer Reliability Analysis* (CRA) approach to propagate the effects of faults and to apply reliability increasing techniques from lower levels of abstraction up to the system level. An important concept of this approach is that for each relevant error model at a specific design level of abstraction, an appropriate reliability analysis shall be applicable and seamlessly integrated into CRA. To realize a holistic cross-layer analysis, the CRA features three important aspects: (a) Individual analysis techniques are encapsulated in *Compositional Reliability Nodes* (CRNs) at *reliability levels of abstraction*, (b) *composition* and *decomposition* are applied to tame system complexity, and (c) formerly incompatible reliability levels of abstraction and, hence, analysis techniques are connected by *adapters*. As a case study, the work in [3] is applied to a concrete reliability analysis during ESL design space exploration of an MPSoC system where thermal-related aging at gate level is the premier reliability threat.

In [5], a cross-layer soft error rate estimation platform which models the error generation and propagation from device level up to application level is presented. It computes soft error vulnerability of all components (gates, flip-flops, and memory arrays) with respect to the running workload on an MPSoC. This is achieved using a SER analysis tool which combines device-level TCAD simulation with cell-level SPICE simulations [6], an analytical circuit-level error propagation method [7] augmented with an FPGA-based approach for fast and rigorous fault injection [8].

The work in hand integrates these approaches and provides a general framework for concurrent consideration of the effects of all fault types, e.g., permanent and transient faults. This framework exploits CRNs and adapters in a tree-based fashion to a flexible and holistic system-wide analysis. The proposed framework also uses the concepts of decomposition and composition similar to the work in [3] (shown in Fig. 1). Due to the complexities, the system is not considered as a whole at lower levels, but, is decomposed into some components (sub-systems). To apply system-level measures at low levels, adapters are employed between adjacent reliability levels. They transform output measures provided at a higher level to input measures required at the lower level of abstraction. Then, necessary analysis approaches are performed at low reliability levels for each component separately. Finally, the output measures provided at the lower levels are transformed to the required input measures at the higher levels using the adapters.

The proposed approach enhances lifetime prediction of a system by concurrent analysis of the effects of different fault types and origins. As a concrete example, we employ the following analysis techniques in our framework in order to combine the analysis of transient and permanent faults at system-level:

- Temperature simulation using HotSpot tool [9] for permanent fault analysis. HotSpot requires a power trace of each component and delivers a temperature profile for the respective component.
- Soft error rate estimation for transient fault analysis [5] using fault injection
- Reliability analysis using Success Trees for a combination of both fault types at system level

REFERENCES

[1] A. Dixit and A. Wood. The impact of new technology on soft error rates. In *International Reliability Physics Symposium*, pages 5B–4, 2011.
[2] T. Streichert et al. Design space exploration of reliable networked embedded systems. *Journal of Systems Architecture: the EUROMICRO Journal*, 53(10):751–763, 2007.
[3] M. Glaß et al. Cross-level compositional reliability analysis for embedded systems. In *Proceedings of the 31st International Conference on Computer Safety, Reliability and Security (SAFECOMP 2012)*, pages 111–124, Magdeburg, Germany, 2012.
[4] Andreas Herkersdorf et. al. Resilience articulation point (rap): Cross-layer dependability modeling for nanometer system-on-chip resilience. *Microelectronics Reliability*, 2014.
[5] M. Ebrahimi et al. Comprehensive Analysis of Alpha and Neutron Particle-induced Soft Errors in an Embedded Processor at Nanoscales. In *Design, Automation and Test in Europe Conference*, pages 1–6, 2014.
[6] E. Costenaro et al. A practical approach to single event transient analysis for highly complex design. *Journal of Electronic Testing*, 29(3):301–315, 2013.
[7] L. Chen et al. Cep: Correlated error propagation for hierarchical soft error analysis. *Journal of Electronic Testing*, 29(2):143–158, 2013.
[8] M. Ebrahimi et al. A fast, flexible, and easy-to-develop fpga-based fault injection technique. *Microelectronics Reliability*, 2014.
[9] K. Skadron et al. Temperature-aware microarchitecture. In *Proceedings of the 30th Annual International Symposium on Computer Architecture*, pages 2–13, 2003.